\documentclass[a4paper,onecolumn,11pt,accepted=2023-11-23]{quantumarticle}
\pdfoutput=1
\usepackage[utf8]{inputenc}
\usepackage[english]{babel}
\usepackage[T1]{fontenc}
\usepackage{amsmath}
\usepackage[colorlinks=true,citecolor=blue]{hyperref}
\usepackage{subfigure}
\usepackage[sort&compress,numbers]{natbib}
\usepackage{tikz}
\usepackage{lipsum}

\begin{document}

\title{Improving the performance of twin-field quantum key distribution with advantage distillation technology}

\author{Hong-Wei Li}
\email{lihow@ustc.edu.cn}
\affiliation{Henan Key Laboratory of Quantum Information and Cryptography, SSF IEU, Zhengzhou 450000, China}
\author{Rui-Qiang Wang}
\affiliation{CAS Key Laboratory of Quantum Information, University of Science and Technology of China, Hefei, Anhui 230026, China}
\author{Chun-Mei Zhang}
\affiliation{Institute of Quantum Information and Technology, Nanjing University of Posts and Telecommunications, Nanjing 210003, China}
\author{Qing-Yu Cai}
\email{qycai@wipm.ac.cn}
\affiliation{School of Information and Communication Engineering, Hainan University, Haikou 570228, China}
\maketitle

\begin{abstract}
In this work, we apply the advantage distillation method to improve the performance of a practical twin-field quantum key distribution system under collective attack.  Compared with the previous analysis result given by Maeda, Sasaki and Koashi [Nature Communication 10, 3140 (2019)], the maximal transmission distance obtained by our analysis method will be increased from 420 km to 470 km. By increasing the loss-independent misalignment error to $12\%$, the previous analysis method can not overcome the rate-distance bound. However, our analysis method can still overcome the rate-distance bound when the misalignment error is $16\%$. More surprisingly, we prove that twin-field quantum key distribution can generate positive secure key even if the misalignment error is close to $50\%$, thus our analysis method can significantly improve the performance of a practical twin-field quantum key distribution system.
\end{abstract}

\section{\label{sec_intro}Introduction}
Quantum key distribution (QKD) \cite{bb84} is the art of sharing information-theoretical secure key between two different remote parties Alice and Bob. Under the perfect quantum devices preparation, the eavesdropper Eve can not get the secure key information even if she has unlimited computation and storage power \cite{sec1, sec2, renner2008security}. Unfortunately, a practical QKD system is usually composed of imperfect devices, and the practical QKD system may be attacked \cite{attck1,attck2,attck3,attck4} by utilizing imperfect quantum state preparation and measurement devices. To avoid the detector side channel attack \cite{attck2,attck3}, measurement-device-independent QKD (MDI-QKD) was proposed  \cite{braunstein2012side,lo2012measurement}. In MDI-QKD, the ideal quantum states are randomly prepared by Alice and Bob, and then will be transmitted to untrusted Charlie to apply Bell state measurement \cite{kwek2021chip}. To beat the Pirandola-Laurenza-Ottaviani-Banchi (PLOB) bound \cite{PLOB}, twin-field QKD (TF-QKD) \cite{TFQKD1} was proposed which is a variant of the MDI-QKD protocol. TF-QKD can overcome the rate scaling from $\eta$ to $\sqrt{\eta}$ with a relatively simple setup, where $\eta$
is the single-photon transmissivity of the link from Alice to Bob. In the TF-QKD protocol, Charlie simply conducts an interference measurement to learn the relative phase between Alice and Bob, and the secure key can be generated with in-phase and anti-phase measurement outcomes respectively. More recently, many intensive studies have been devoted to achieving
information theoretic proofs of variants of TF-QKD protocols \cite{TFQKD2,TFQKD3,TFQKD4,TFQKD5,TFQKD6,TFQKD7,curras,curras2021twin}, and several practical TF-QKD systems have been widely implemented in labs and field tests \cite{TFQKDEX1,TFQKDEX2,TFQKDEX3,TFQKDEX4,TFQKDEX5,TFQKDEX6,TFQKDEX7}. By applying the entanglement purification with two-way classical communication, the transmission distance of TF-QKD protocol can be improved \cite{TFQKD8}.

To analyze the security of MDI-QKD protocols \cite{braunstein2012side,lo2012measurement}, Alice and Bob should randomly prepare phase-randomized pulses in Z basis and X basis respectively, where Z basis consists of $|0\rangle$
and $|1\rangle$, and X basis consists of $|+\rangle=\frac{|0\rangle+|1\rangle}{\sqrt{2}}$
and $|-\rangle=\frac{|0\rangle-|1\rangle}{\sqrt{2}}$. Note that the bit error rate in Z basis and X basis can also be applied to characterize the bit error rate and phase error rate about the quantum channel. Based on this quantum channel characterization, the upper bound of the secure key information leaked to Eve can be estimated. However, to estimate the bit error rate in X basis, TF-QKD needs to generate a non-classical optical state \cite{TFQKD7,TFQKD8}, which is difficult to realize in current technology. Fortunately, Maeda, Sasaki and Koashi proposed the operator dominance method \cite{TFQKD7} to estimate the bit error rate in X basis by preparing phase-randomized weak coherent states. Combining the error rate in Z basis with the error rate in X basis, the practical quantum channel in TF-QKD can be characterized, and the secure key rate can be analyzed with the entanglement distillation and purification method \cite{sec1, sec2, TFQKD7}. The purpose of TF-QKD is to extend the transmission distance with current technology, but the final solution is to use quantum repeater in the future research \cite{mastriani2020satellite, hu2021long,long2022evolutionary}.

In this work, we apply the error rate in X basis and Z basis to characterize the practical quantum channel. By applying the information-theoretical security analysis method, we apply the advantage distillation (AD) method to improve the secure key rate and transmission distance. The AD method was initially proposed in classical cryptography theory \cite{maurer}, then it has been widely used in different QKD protocols \cite{kraus2007security,bae2007key,murta2020key,tan2020advantage} to improve the error tolerance. More recently, we analyze the security of practical BB84-QKD  \cite{bb84}, six-state-QKD \cite{sixstate} and MDI-QKD  \cite{braunstein2012side,lo2012measurement} systems by combining the AD method \cite{renner2008security} with the decoy-state method  \cite{hwang2003quantum, wang2005beating, lo2005decoy}, and the analysis results demonstrate that the AD method can significantly improve the performance of different practical QKD systems  \cite{li2021improving}.
Inspired by our previous work, we combine the AD method and the operator dominance method  \cite{TFQKD7} to analyze the security of TF-QKD, and the analysis results demonstrate that both the transmission distance and the secure key rate can be sharply improved. More surprisingly, the analysis results also demonstrate that TF-QKD can generate positive secure key even if the loss-independent misalignment error is close to $50\%$, thus our analysis method can significantly improve the robustness of the practical TF-QKD system.

\section{\label{sec_prot}Security of QKD with AD}

To prove the security of a QKD protocol, the bit error rate of different bases should be precisely estimated. It has been proved that the security of the state preparation and measurement based QKD protocol can be analyzed with the entanglement based QKD protocol  \cite{sec1, renner2008security}. In the entanglement based protocol, Alice and Bob can take inputs from four-dimensional Hilbert spaces $H_A\otimes H_B$ to
apply Z basis and X basis measurements. By considering BB84 \cite{bb84} and six-state \cite{sixstate} QKD protocols, it has been proved that
Eve's general attack can be reduced to the Pauli attack \cite{renner2008security, two1}, which can be described by the classical probability theory.
Thus, the QKD protocol can be illustrated with the following quantum state preparation,
\begin{equation}
{\sigma _{AB}} = \sum\limits_{i = 0}^3 {{\lambda _i}} \left| {{\Phi _i}} \right\rangle \left\langle {{\Phi _i}} \right|,with\sum\limits_{i = 0}^3 {{\lambda _i}}  = 1,
\end{equation}
where
\begin{equation}
\begin{array}{lll}
|\Phi_0\rangle=\frac{1}{\sqrt{2}}(|00\rangle+|11\rangle)\\
|\Phi_1\rangle=\frac{1}{\sqrt{2}}(|00\rangle-|11\rangle)\\
|\Phi_2\rangle=\frac{1}{\sqrt{2}}(|01\rangle+|10\rangle)\\
|\Phi_3\rangle=\frac{1}{\sqrt{2}}(|01\rangle-|10\rangle).\\
\end{array}
\end{equation}
By combining this state preparation with the $Z$ basis and $X$ basis  measurement in Alice and Bob's side, the secure key rate can be given by \cite{renner2008security}
\begin{equation}
\begin{aligned}
R \geq &min_{\lambda_0,\lambda_1,\lambda_2,\lambda_3} [S(A|E)-H(A|B)] \\
= &min_{\lambda_0,\lambda_1,\lambda_2,\lambda_3}  [ 1-(\lambda_0+\lambda_1)H(\frac{\lambda_0}{\lambda_0+\lambda_1})\\
&-(\lambda_2+\lambda_3)H(\frac{\lambda_2}{\lambda_2+\lambda_3})-H(\lambda_0+\lambda_1)],
\end{aligned}
\end{equation}
where $E$ is Eve's ancillary state, $S(A|E)=S(A,E)-S(E)$, $H(A|B)=H(A,B)-H(B)$, $H(x)=-x\text {log}x-(1-x)\text {log}(1-x)$ and $S(\rho)=-\text {tr}(\rho \text {log}\rho)$ are entropy functions. Since the quantum channel can be controlled by Eve, she can choose the optimal parameters $\lambda_i,~i=\{0,1,2,3\}$ to reduce the secure key rate, but $\lambda_i$ should also be restricted by the quantum bit error rate in two different bases.

To improve the maximal tolerable error rate, the repetition code protocol based AD method has been proposed \cite{renner2008security}. In the repetition code protocol, Alice and Bob split their raw key into blocks of $b$ bits $x_0,x_1,...x_{b-1}$ and $y_0,y_1,...y_{b-1}$ respectively.
Alice privately generates a random bit $c \in \{0, 1\}$, and sends
the message $m = m_0, m_1, ..., m_{b-1}=x_0\oplus c,x_1\oplus c,...x_{b-1}\oplus c$ to Bob through an
authenticated classical channel. Bob accepts the block if and only if $m_0\oplus y_0, m_1\oplus y_1, ..., m_{b-1}\oplus y_{b-1} \in \{0, 0, ... 0~\text {or}~1, 1, ...,1\}$. If Alice and Bob accept the block, they keep the first bit $x_0$ and $y_0$ as the raw key. Finally, Alice and Bob will apply the error correction and privacy amplification algorithms to generate the final secure key.

Based on the repetition code protocol, the secure key rate $\tilde{R}$ can be modified as the following inequality  \cite{renner2008security}
\begin{equation}
\begin{aligned}
\tilde{R} \geq  &max_b~min_{\lambda_0,\lambda_1,\lambda_2,\lambda_3} \frac{1}{b} p_{succ} [1-(\tilde{\lambda}_0+\tilde{\lambda}_1)H(\frac{\tilde{\lambda}_0}{\tilde{\lambda}_0+\tilde{\lambda}_1})\\
&-(\tilde{\lambda}_2+\tilde{\lambda}_3)H(\frac{\tilde{\lambda}_2}{\tilde{\lambda}_2+\tilde{\lambda}_3})-H(\tilde{\lambda}_0+\tilde{\lambda}_1)],
\end{aligned}
\end{equation}

where
\begin{equation}
\begin{aligned}
\tilde{\lambda}_0=\frac{(\lambda_0+\lambda_1)^b+(\lambda_0-\lambda_1)^b}{2p_{succ}},\\
\tilde{\lambda}_1=\frac{(\lambda_0+\lambda_1)^b-(\lambda_0-\lambda_1)^b}{2p_{succ}},\\
\tilde{\lambda}_2=\frac{(\lambda_2+\lambda_3)^b+(\lambda_2-\lambda_3)^b}{2p_{succ}},\\
\tilde{\lambda}_3=\frac{(\lambda_2+\lambda_3)^b-(\lambda_2-\lambda_3)^b}{2p_{succ}},
\end{aligned}
\end{equation}
$p_{succ}=(\lambda_0+\lambda_1)^b+(\lambda_2+\lambda_3)^b$ is the successful probability of the AD method. Since the parameter $b$ in the AD method can be controlled by Alice and Bob, they can choose the optimal $b$ to improve the secure key rate. Note that this secure key rate is based on the single photon state preparation, which can also be applied in practical QKD systems with weak coherent pulse preparation \cite{li2021improving}.

\section{\label{sec_prot}The virtual protocol to analyze the quantum channel}

Based on the TF-QKD protocol proposed by Maeda, Sasaki and Koashi \cite{TFQKD7}, Alice and Bob generate four different pulses with the signal state modulation and the testing state modulation respectively. In the signal state modulation, Alice and Bob randomly prepare the weak coherent pulse with amplitude $\sqrt{\mu}$ or $-\sqrt{\mu}$. In the testing state modulation, Alice and Bob randomly prepare the phase-randomized weak coherent pulse with intensities $\nu_1$, $\nu_2$ and $0$ respectively. For every pair of pulses received from Alice and Bob, Charlie
announces whether the phase difference is successfully detected. When the phase difference is detected, Charlie further announces whether it is in-phase when detector $D_1$ clicks or anti-phase
when detector $D_2$ clicks.  After receiving Charlie's measurement outcomes, Alice and Bob will apply phase information on the signal state to generate the sifted key. More precisely, Charlie's measurement outcomes can be divided into two cases. In the in-phase measurement outcome case, Alice and Bob will respectively generate the random bit $a$ with the signal state preparation $|(-1)^a\sqrt{\mu}\rangle$. In the anti-phase measurement outcome case, Alice and Bob will respectively generate the random bit $a$ with the signal state preparation $|(-1)^a\sqrt{\mu}\rangle$ and $|(-1)^{a\oplus1}\sqrt{\mu}\rangle$ respectively.

By applying the entanglement based protocol, the signal state preparation in Alice's side can be illustrated with the following quantum state
\begin{equation}
\begin{array}{lll}
\frac{|0\rangle_A|\sqrt{\mu}\rangle_{C_A}+|1\rangle_A|-\sqrt{\mu}\rangle_{C_A}}{\sqrt{2}}.
\end{array}
\end{equation}
After preparing the quantum states $|\psi\rangle_{AC_A}$, Alice will measure the first quantum state, and the second quantum state will be transmitted to the quantum channel. And the signal state preparation in Bob's side can be illustrated with the following quantum state.
\begin{equation}
\begin{array}{lll}
\frac{|0\rangle_B|\sqrt{\mu}\rangle_{C_B}-|1\rangle_B|-\sqrt{\mu}\rangle_{C_B}}{\sqrt{2}}.
\end{array}
\end{equation}
After preparing the quantum states $|\varphi\rangle_{BC_B}$, Bob will measure the first quantum state, and the second quantum state will be transmitted to the quantum channel.

It should be pointed out that this analysis method has been given in Ref. \cite{TFQKD7}. Maeda, Sasaki and Koashi have proved that Alice and Bob's procedure in the signal mode can be equivalently executed by preparing the qubits $AB$ and the optical pulses $C_AC_B$ in a joint quantum state
$\frac{|0\rangle_A|\sqrt{\mu}\rangle_{C_A}+|1\rangle_A|-\sqrt{\mu}\rangle_{C_A}}{\sqrt{2}}\otimes\frac{|0\rangle_B|\sqrt{\mu}\rangle_{C_B}-|1\rangle_B|-\sqrt{\mu}\rangle_{C_B}}{\sqrt{2}}$. After preparing these quantum states, Alice and Bob will measure the first quantum state, and the second quantum state will be transmitted to Charlie.
Thus the quantum state shared among Alice, Bob and Charlie can be given by
\begin{equation}
\begin{array}{lll}
&\frac{|0\rangle_A|\sqrt{\mu}\rangle_{C_A}+|1\rangle_A|-\sqrt{\mu}\rangle_{C_A}}{\sqrt{2}}\otimes\frac{|0\rangle_B|\sqrt{\mu}\rangle_{C_B}-|1\rangle_B|-\sqrt{\mu}\rangle_{C_B}}{\sqrt{2}}\\
=& \frac{1}{4}[(|0\rangle|0\rangle+|1\rangle|1\rangle)_{AB}(|\sqrt{\mu}\rangle|\sqrt{\mu}\rangle-|-\sqrt{\mu}\rangle|-\sqrt{\mu}\rangle)_{C_A, C_B}\\
&+(|0\rangle|0\rangle-|1\rangle|1\rangle)_{AB}(|\sqrt{\mu}\rangle|\sqrt{\mu}\rangle+|-\sqrt{\mu}\rangle|-\sqrt{\mu}\rangle)_{C_A, C_B}\\
&+(|0\rangle|1\rangle+|1\rangle|0\rangle)_{AB}(-|\sqrt{\mu}\rangle|-\sqrt{\mu}\rangle+|-\sqrt{\mu}\rangle|\sqrt{\mu}\rangle)_{C_A, C_B}\\
&+(|0\rangle|1\rangle-|1\rangle|0\rangle)_{AB}(-|\sqrt{\mu}\rangle|-\sqrt{\mu}\rangle-|-\sqrt{\mu}\rangle|\sqrt{\mu}\rangle)_{C_A, C_B}].
\end{array}
\end{equation}
After measuring the received quantum states $C_A$ and $C_B$  with a
$50:50$ beam-splitter and a pair of photon detectors $D_1$ and $D_2$, Charlie announces whether the phase difference is successfully detected.
Note that TF-QKD protocol is based on the first-order interference, while MDI-QKD is based on the second-order interference. Correspondingly,
this quantum state will be transformed to
\begin{equation}
\begin{array}{lll}
\frac{1}{4}[(|0\rangle|0\rangle+|1\rangle|1\rangle)_{AB}(|\sqrt{2\mu}\rangle|0\rangle-|-\sqrt{2\mu}\rangle|0\rangle)_{D_1, D_2}+\\
(|0\rangle|0\rangle-|1\rangle|1\rangle)_{AB}(|\sqrt{2\mu}\rangle|0\rangle+|-\sqrt{2\mu}\rangle|0\rangle)_{D_1, D_2}+\\
(|0\rangle|1\rangle+|1\rangle|0\rangle)_{AB}(-|0\rangle|\sqrt{2\mu}\rangle+|0\rangle|-\sqrt{2\mu}\rangle)_{D_1, D_2}+\\
(|0\rangle|1\rangle-|1\rangle|0\rangle)_{AB}(-|0\rangle|\sqrt{2\mu}\rangle-|0\rangle|-\sqrt{2\mu}\rangle)_{D_1, D_2}],
\end{array}
\end{equation}
where the detection results that detector $D_1$ clicks and detector $D_2$ clicks respectively demonstrate the in-phase and anti-phase measurement outcomes in Charlie's side.
Here, for the simplicity of the discussion, we assume that there are no channel losses. Correspondingly, the quantum states $(|\sqrt{2\mu}\rangle|0\rangle-|-\sqrt{2\mu}\rangle|0\rangle)_{D_1, D_2}$ and $(|\sqrt{2\mu}\rangle|0\rangle+|-\sqrt{2\mu}\rangle|0\rangle)_{D_1, D_2}$ demonstrate the in-phase measurement outcome, while quantum states $(-|0\rangle|\sqrt{2\mu}\rangle+|0\rangle|-\sqrt{2\mu}\rangle)_{D_1, D_2}$ and $(-|0\rangle|\sqrt{2\mu}\rangle-|0\rangle|-\sqrt{2\mu}\rangle)_{D_1, D_2}$ demonstrate the anti-phase measurement outcome.  Note that TF-QKD protocol is a variant of the MDI-QKD protocol, thus the measurement outcomes can be assumed to be controlled by Eve. We can simply assume the click of detector $D_1$ demonstrates the Bell state $\frac{1}{\sqrt{2}}(|0\rangle|0\rangle+|1\rangle|1\rangle)_{AB}$ preparation in Alice and Bob's side, while the click of detector $D_2$ demonstrates the Bell state $\frac{1}{\sqrt{2}}(|0\rangle|1\rangle+|1\rangle|0\rangle)_{AB}$ preparation in Alice and Bob's side. Since Charlie's measurement outcomes have no security requirement, this assumption is reasonable, and more detailed explanation has been given in Ref. \cite{TFQKD7}.

By applying the time reversed entanglement technique, we can assume the Bell states $\frac{1}{\sqrt{2}}(|0\rangle|0\rangle+|1\rangle|1\rangle)_{AB}$ and $\frac{1}{\sqrt{2}}(|0\rangle|1\rangle+|1\rangle|0\rangle)_{AB}$ is prepared in Charlie's side, then the two quantum states will be transmitted to Alice and Bob to perform Z basis or
X basis measurement. Based on this analysis method, the in-phase measurement outcome demonstrates the Bell state preparation
\begin{equation}
\begin{array}{lll}
\frac{1}{\sqrt{2}}(|0\rangle|0\rangle+|1\rangle|1\rangle)_{AB}
=\frac{1}{\sqrt{2}}(|+\rangle|+\rangle+|-\rangle|-\rangle)_{AB}.
\end{array}
\end{equation}
Suppose Alice and Bob apply Z basis measurement, Z basis error is defined to be an event where the pair is found in either state $|0\rangle|1\rangle_{AB}$ or  $|1\rangle|0\rangle_{AB}$. Suppose Alice and Bob apply X basis measurement, X basis error is defined to be an event where the pair is found in either state $|+\rangle|-\rangle_{AB}$ or  $|-\rangle|+\rangle_{AB}$.

Similarly, the anti-phase measurement outcome demonstrates the Bell state preparation
\begin{equation}
\begin{array}{lll}
\frac{1}{\sqrt{2}}(|0\rangle|1\rangle+|1\rangle|0\rangle)_{AB}=\frac{1}{\sqrt{2}}(|+\rangle|+\rangle-|-\rangle|-\rangle)_{AB}.
\end{array}
\end{equation}
Suppose Alice and Bob apply Z basis measurement, Z basis error is defined to be an event where the pair was found in either state $|0\rangle|0\rangle_{AB}$ or  $|1\rangle|1\rangle_{AB}$. Suppose Alice and Bob apply X basis measurement, X basis error is defined to be an event where the pair is found in either state $|+\rangle|-\rangle_{AB}$ or  $|-\rangle|+\rangle_{AB}$.

In a practical TF-QKD experiment, the bit error rate in Z basis can be directly tested, but the bit error rate in X basis can't be directly observed.
To analyze the bit error rate in X basis, the quantum state shared among Alice, Bob and Charlie can be rewritten as
\begin{equation}
\begin{array}{lll}
&\frac{|0\rangle_A|\sqrt{\mu}\rangle_{C_A}+|1\rangle_A|-\sqrt{\mu}\rangle_{C_A}}{\sqrt{2}}\otimes\frac{|0\rangle_B|\sqrt{\mu}\rangle_{C_B}-|1\rangle_B|-\sqrt{\mu}\rangle_{C_B}}{\sqrt{2}}\\
=&(\sqrt{c_+}|+\rangle_A|\sqrt{\mu_{even}}\rangle_{C_A}+\sqrt{c_-}|-\rangle_A|\sqrt{\mu_{odd}}\rangle_{C_A})\\
&\otimes(\sqrt{c_-}|+\rangle_B|\sqrt{\mu_{odd}}\rangle_{C_B}+\sqrt{c_+}|-\rangle_B|\sqrt{\mu_{even}}\rangle_{C_B}),
\end{array}
\end{equation}
where $c_+:=e^{-\mu}cosh\mu$, $c_-:=e^{-\mu}sinh\mu$, $|\sqrt{\mu_{even}}\rangle=(|\sqrt{\mu}\rangle+|-\sqrt{\mu}\rangle)/(2\sqrt{c_+})$ and $|\sqrt{\mu_{odd}}\rangle=(|\sqrt{\mu}\rangle-|-\sqrt{\mu}\rangle)/(2\sqrt{c_-})$. Based on this state preparation, the X basis error occurs with probability $p_{even}=c_+^2+c_-^2=e^{-2\mu}cosh2\mu$, and the optical pulses are sent in the following quantum state
\begin{equation}
\begin{array}{lll}
p_{even}\rho^{even}=&c_+^2|\sqrt{\mu_{even}}\sqrt{\mu_{even}}\rangle\langle\sqrt{\mu_{even}}\sqrt{\mu_{even}}|_{C_AC_B}\\
&+c_-^2|\sqrt{\mu_{odd}}\sqrt{\mu_{odd}}\rangle\langle\sqrt{\mu_{odd}}\sqrt{\mu_{odd}}|_{C_AC_B}.
\end{array}
\end{equation}
To analyze the error rate in X basis, the counting rate with this non-classical optical state preparation can be analyzed with the operator dominance method \cite{TFQKD7}, where the detection frequency of $\rho^{even}$ can be estimated
from a combination of phase-randomized weak coherent states with intensities $\nu_1$, $\nu_2$ and $0$. Combining this analysis result with the  operator dominance method given by \cite{TFQKD7}, we can estimate the quantum channel parameters $\lambda_i,~i=\{0,1,2,3\}$ in the following section.

\section{\label{sec_prot}Security of TF-QKD with AD}

By applying the previous analysis result with the signal state modulation in Alice and Bob's side, we need to analyze the error rate $E_{uu}^{ZZ}$ in Z basis and the error rate $E_{uu}^{XX}$ in X basis respectively.
In a practical TF-QKD experiment, $E_{uu}^{ZZ}$ can be directly calculated by testing part of the in-phase and anti-phase measurement outcomes. However, the non-classical optical state $\rho^{even}$ is hard to realize in current technology, the bit error rate $E_{uu}^{XX}$  can not be directly estimated in a practical TF-QKD system.
Fortunately, Maeda, Sasaki and Koashi proposed the operator dominance method to estimate $E_{uu}^{XX}$ \cite{TFQKD7}, where the linear combination of the testing states can be applied to approximate the non-classical optical state $\rho^{even}$. To analyze the security of the entanglement based TF-QKD protocol with the information-theoretical analysis method, the relationship among  $E_{uu}^{ZZ}$, $E_{uu}^{XX}$ and $\lambda_i$  can be given with the following equations,
\begin{equation}
\begin{aligned}
&\lambda_1+\lambda_3=E_{uu}^{XX}, \\
&\lambda_2+\lambda_3=E_{uu}^{ZZ},
\end{aligned}
\end{equation}
where $\lambda_0+\lambda_1+\lambda_2+\lambda_3=1$. In the TF-QKD protocol, only the signal state can be used to generate the final secure key. Since the quantum channel parameters $\lambda_i,~i=\{0,1,2,3\}$ can be estimated with Eq. (14),
Eq. (4) can be modified with the following inequality by applying the AD method
\begin{equation}\label{GLLP}
\begin{aligned}
R_{TF} \geq &max_b~ \frac{1}{b} q_{succ}^{ZZ} Q_{uu}^{ZZ} [S(A|E)-H(A|B)] \\
= &max_b~\frac{1}{b} q_{succ}^{ZZ}Q_{uu}^{ZZ} [(1-(\tilde{\lambda}_0+\tilde{\lambda}_1)H(\frac{\tilde{\lambda}_0}{\tilde{\lambda}_0+\tilde{\lambda}_1})\\
&-(\tilde{\lambda}_2+\tilde{\lambda}_3)H(\frac{\tilde{\lambda}_2}{\tilde{\lambda}_2+\tilde{\lambda}_3}))-fh(\tilde{E}_{uu}^{ZZ})],
\end{aligned}
\end{equation}
where $Q_{uu}^{ZZ}$  is the counting rate by considering Alice and Bob prepare the signal states, $\tilde{E}_{uu}^{ZZ}=\frac{{E_{uu}^{ZZ}}^b}{{E_{uu}^{ZZ}}^b+(1-E_{uu}^{ZZ})^b}$ is the error rate after the AD protocol, $f>1$ is the error correction efficiency, $q_{succ}^{ZZ}={E_{uu}^{ZZ}}^b+(1-E_{uu}^{ZZ})^b$ is the successful probability of the AD method in the practical TF-QKD system.

By applying the operator dominance method with asymptotic key length, the error rate in X basis $E_{\mu\mu}^{XX}$  can be given by \cite{TFQKD7}
\begin{equation}
\begin{array}{lll}
E_{\mu\mu}^{XX}=C_1(1+\sqrt{(C_2+C_4)C_3})^2,
\end{array}
\end{equation}
where $C_1=\frac{e^{-2\mu}d}{1-e^{-2\mu\eta}+e^{-2\mu\eta}d}$, $C_2=\frac{e^{-2\nu_1}(\nu_1-\nu_2)}{\mu_2}$, $C_3=\frac{1}{\nu_1e^{-2\nu_1}}\sum_{k=1}^{\infty}\frac{\mu^{2k}(k+1)}{\nu_1^{2k-1}-\nu_2^{2k-1}}$, $C_4=\frac{1}{d}(1-e^{-2\nu_1\eta}+e^{-2\nu_1\eta}d-\frac{\nu_1e^{-2\nu_1}}{\nu_2e^{-2\nu_2}}(1-e^{-2\nu_2\eta}+e^{-2\nu_2\eta}d))$,
$max(\mu,~\nu_2)<\nu_1$. Note that, we assume each
detector has a dark count probability of $p_d$, which amounts to the effective
probability $d= 2p_d- {p_d}^2$ from the two detectors. By considering the channel transmission efficiency, the overall transmissivity from
Alice (Bob) to Charlie's detection is $\eta$.

Based on the simulation parameters given by \cite{TFQKD7}, we calculate the secure key rate $R_{TF}$ as a function of secure key rate transmission distance $L$ between Alice and Bob with different loss-independent misalignment error.  The loss-independent misalignment error is the probability that a photon hits the erroneous detector, which is independent of the transmission distance. In a practical experimental realization,  the loss-independent misalignment error can be applied to characterize the alignment and stability of the optical system. We assume a fiber loss of $0.2$ dB/km, a loss-independent misalignment
error of $e_d=0.03$, error correction efficiency $f=1.1$, each detector has a detection efficiency $\eta_d=0.3$ and dark count probability $p_d=10^{-8}$.
The overall transmissivity from Alice (Bob) to Charlie's detection is $\eta=\eta_d 10^{-0.01L}$. In the asymptotic
limit, the counting rate of Alice and Bob's signal states $Q_{uu}^{ZZ}$ can be given by
\begin{equation}
\begin{array}{lll}
Q_{uu}^{ZZ}=1-e^{-2\mu\eta}+e^{-2\mu\eta}d,\\
\end{array}
\end{equation}
Correspondingly, the error rate in Z basis $E_{\mu\mu}^{ZZ}$ can be given by
\begin{equation}
\begin{array}{lll}
E_{\mu\mu}^{ZZ}=\frac{e_d(1-e^{-2\mu\eta})+\frac{e^{-2\mu\eta}d}{2}}{Q_{uu}^{ZZ}}.\\
\end{array}
\end{equation}

Based on these simulation parameters and the optimal $b$ values, we calculate the secure key rate $R_{TF}$  as a function of transmission distance between Alice and Bob in Figure 1.
\begin{figure*}[!ht]
	\centering
	\subfigure[The relationship between the
	transmission distance and the secure key rate, the blue line is the secure key rate given by \cite{TFQKD7}, the red line is the secure key rate  given by this work, and the green line is the PLOB bound. ]{
		\label{fig:subfig:a}
		\includegraphics[width=3in]{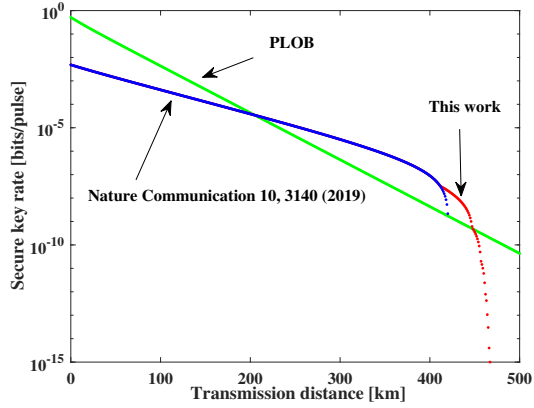}    }
	\subfigure[The optimal $b$ and success probability of advantage distillation with different
	transmission distances.  From $0$ km to $411$ km, the optimal $b$ value is $1$ and the success probability of advantage distillation is $1$, thus we do not need to utilize the AD method in these transmission distances. ]{
		\label{fig:subfig:b}
		\includegraphics[width=3in]{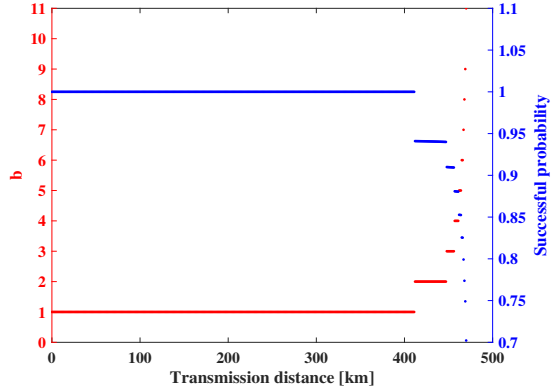}    }
	\caption{Results of TF-QKD protocol with $e_d=0.03$.}    %
	\label{fig:subfig}
\end{figure*}
Comparing with the previous analysis result given by \cite{TFQKD7}, we find that both of the two analysis results can overcome the PLOB bound $-\textrm{log}(1-\eta_d 10^{-0.02L})$, but the maximal secure key transmission distance can be improved from $420$ km to $470$ km.

By increasing the loss-independent misalignment
error to $e_d=0.12$, we calculate the secure key rate in Figure 2.
\begin{figure*}[!ht]
	\centering
	\subfigure[The relationship between the
	transmission distance and the secure key rate, the blue line is the secret key rate given by \cite{TFQKD7}, the red line is the secure key rate  given by this work, and the green line is the PLOB bound. ]{
		\label{fig:subfig:a}
		\includegraphics[width=3in]{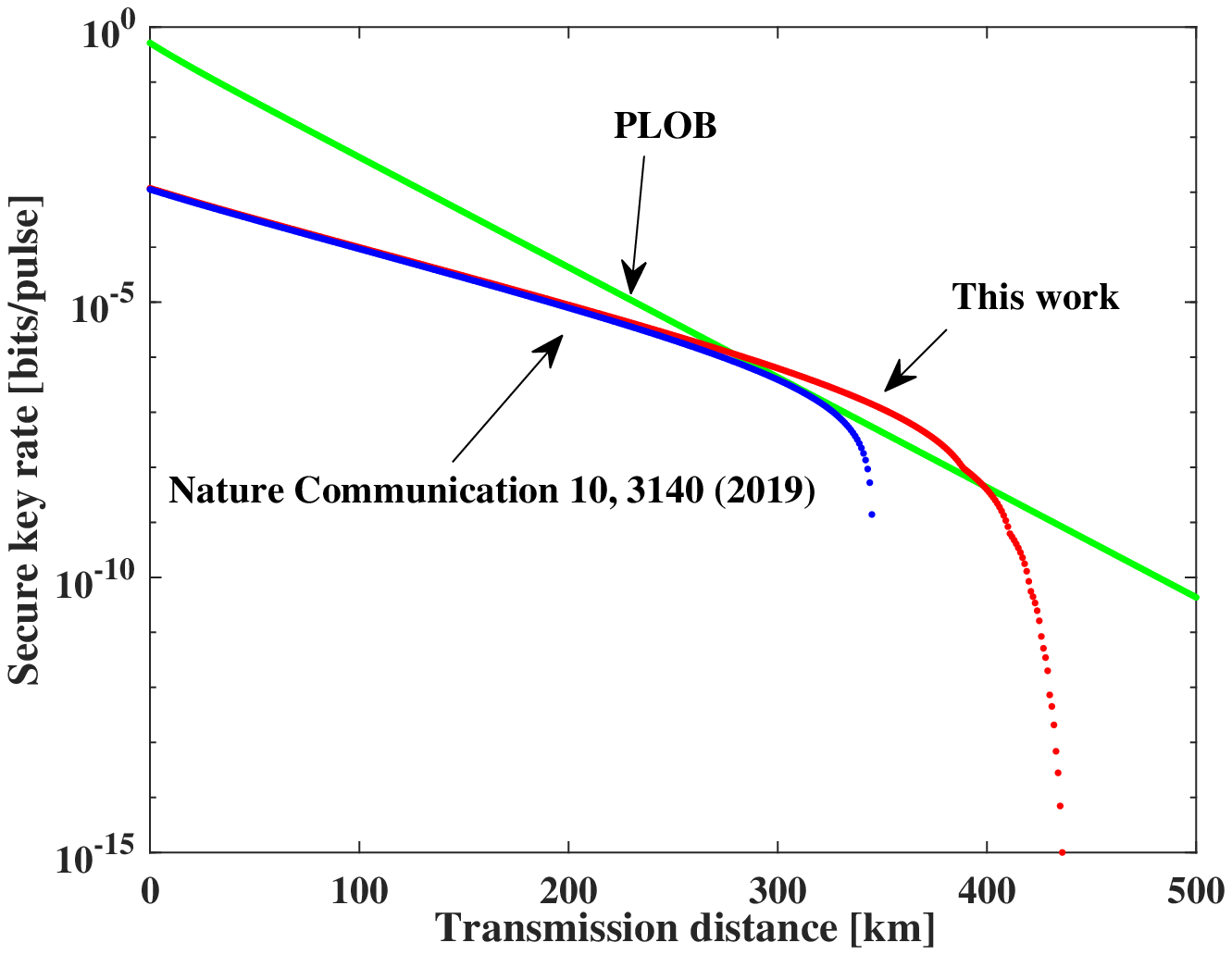}    }
	\subfigure[The relationship between the
	transmission distance and the optimal $b$ values. From $0$ km to $388$ km, the optimal $b$ value is $2$. ]{
		\label{fig:subfig:b}
		\includegraphics[width=3in]{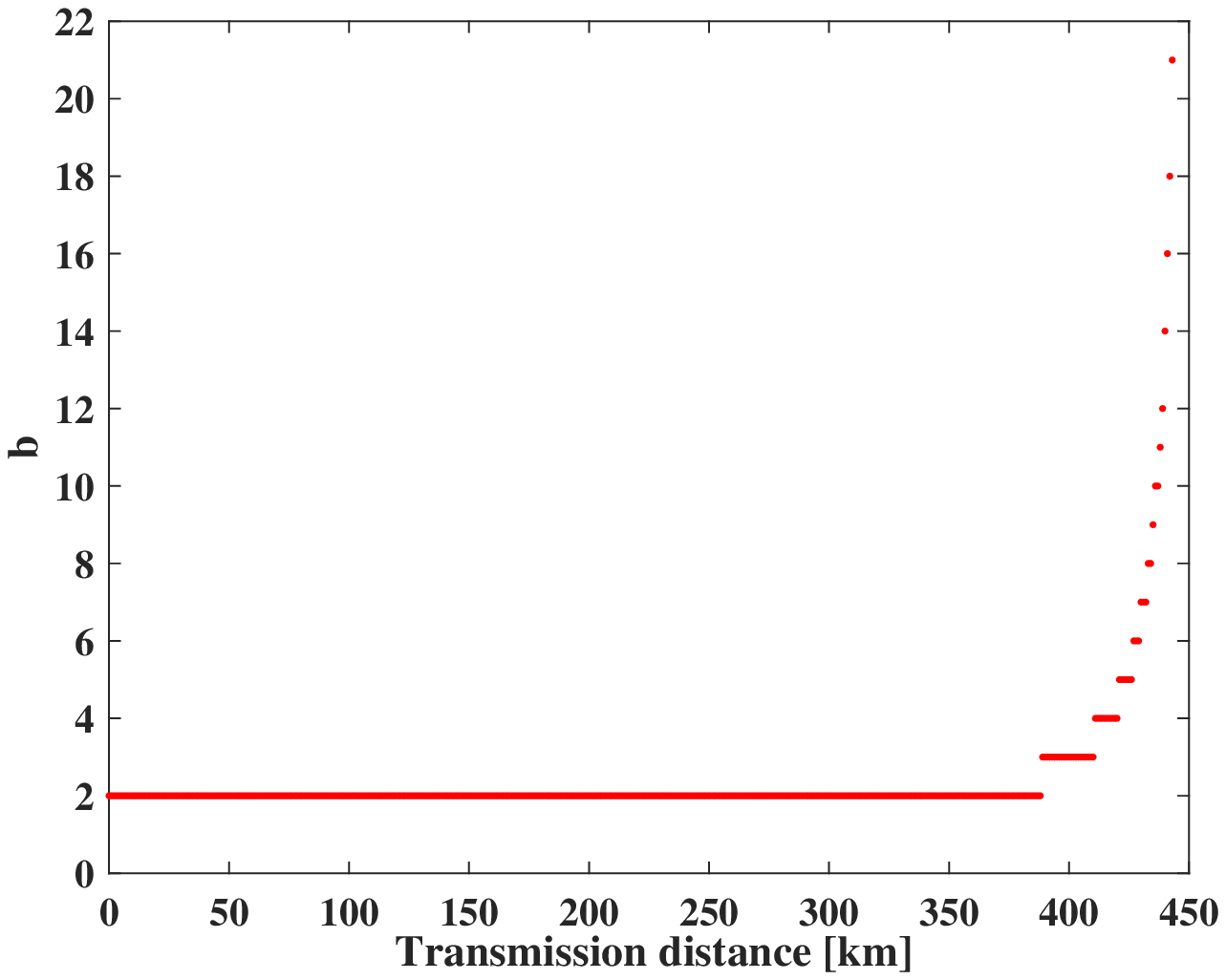}    }
	\caption{Results of TF-QKD protocol with $e_d=0.12$.}    %
	\label{fig:subfig}
\end{figure*}
We find that our analysis result can overcome the PLOB bound and the maximal secure key transmission distance can be improved from $345$ km to $443$ km, but the previous analysis method can not overcome the PLOB bound at any transmission distance.

By increasing the loss-independent misalignment
error to $e_d=0.16$, we calculate the secure key rate in Figure 3.
\begin{figure*}[!ht]
	\centering
	\subfigure[The relationship between the
	transmission distance and the secure key rate, the blue line is the secret key rate given by \cite{TFQKD7}, the red line is the secure key rate  given by this work, and the green line is the PLOB bound. ]{
		\label{fig:subfig:a}
		\includegraphics[width=3in]{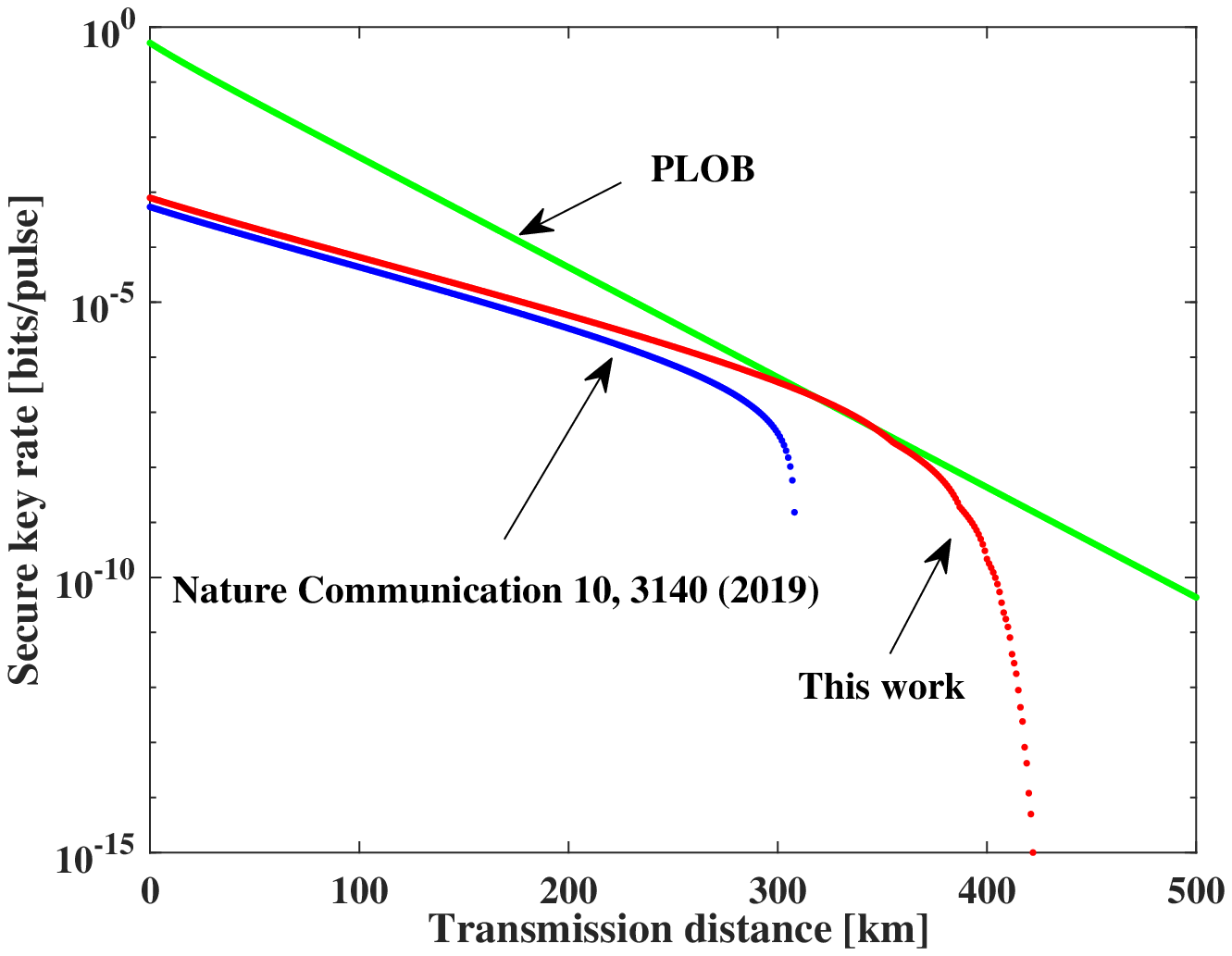}    }
	\subfigure[The relationship between the
	transmission distance and the optimal $b$ values. From $0$ km to $355$ km, the optimal $b$ value is $2$. ]{
		\label{fig:subfig:b}
		\includegraphics[width=3in]{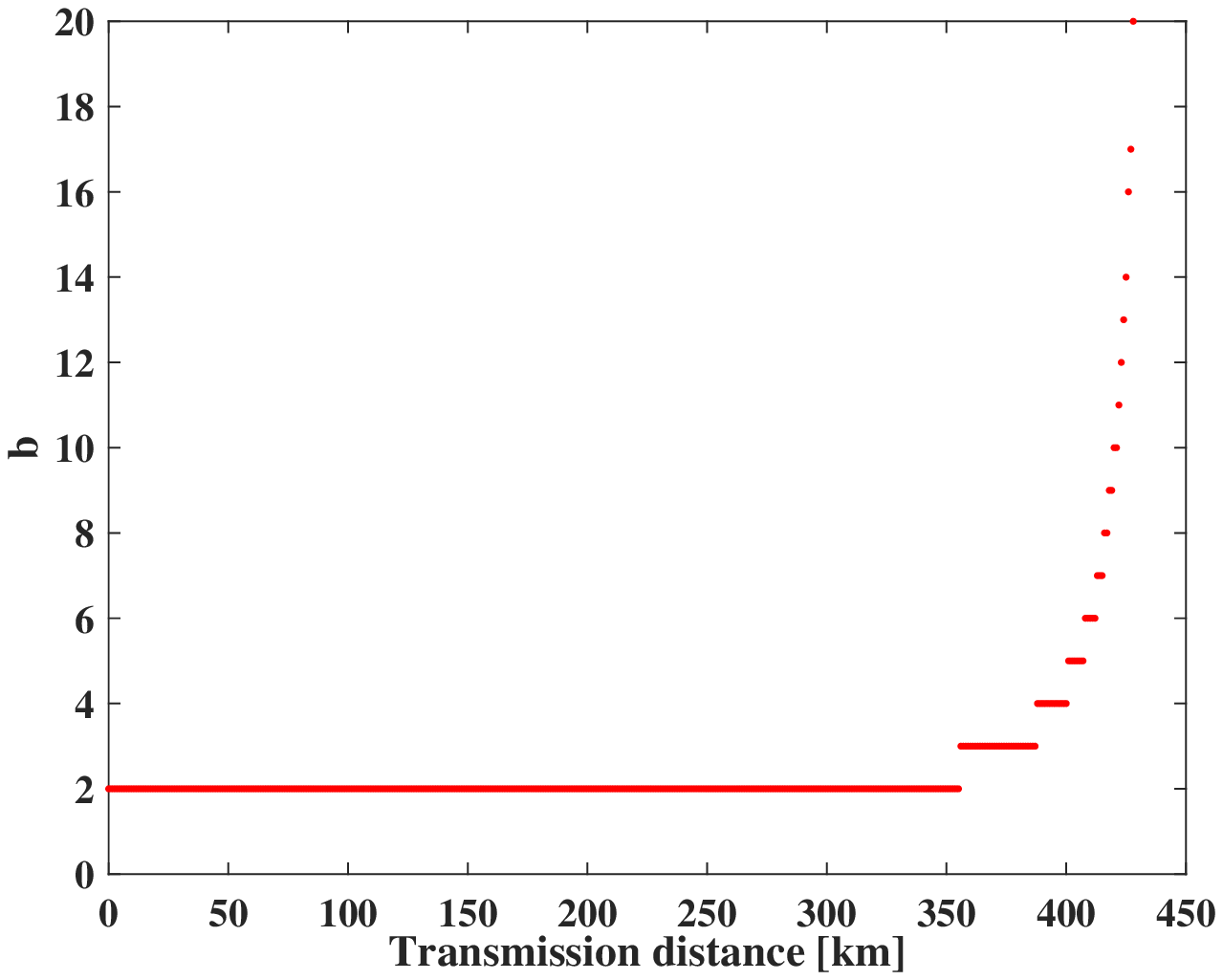}    }
	\caption{Results of TF-QKD protocol with $e_d=0.16$.}    %
	\label{fig:subfig}
\end{figure*}
We find that our analysis result can still overcome the PLOB bound, and the maximal secure key transmission distance can be improved from $308$ km to $428$ km. From the calculation result, we find that the advantage distillation technology has a better effect on improving the secure key rate when the misalignment error rate is high.

By increasing the loss-independent misalignment
error to $e_d=0.48$, we calculate the secure key rate in Figure 4.
\begin{figure*}[!ht]
	\centering
	\subfigure[The relationship between the
	transmission distance and the secure key rate. ]{
		\label{fig:subfig:a}
		\includegraphics[width=3in]{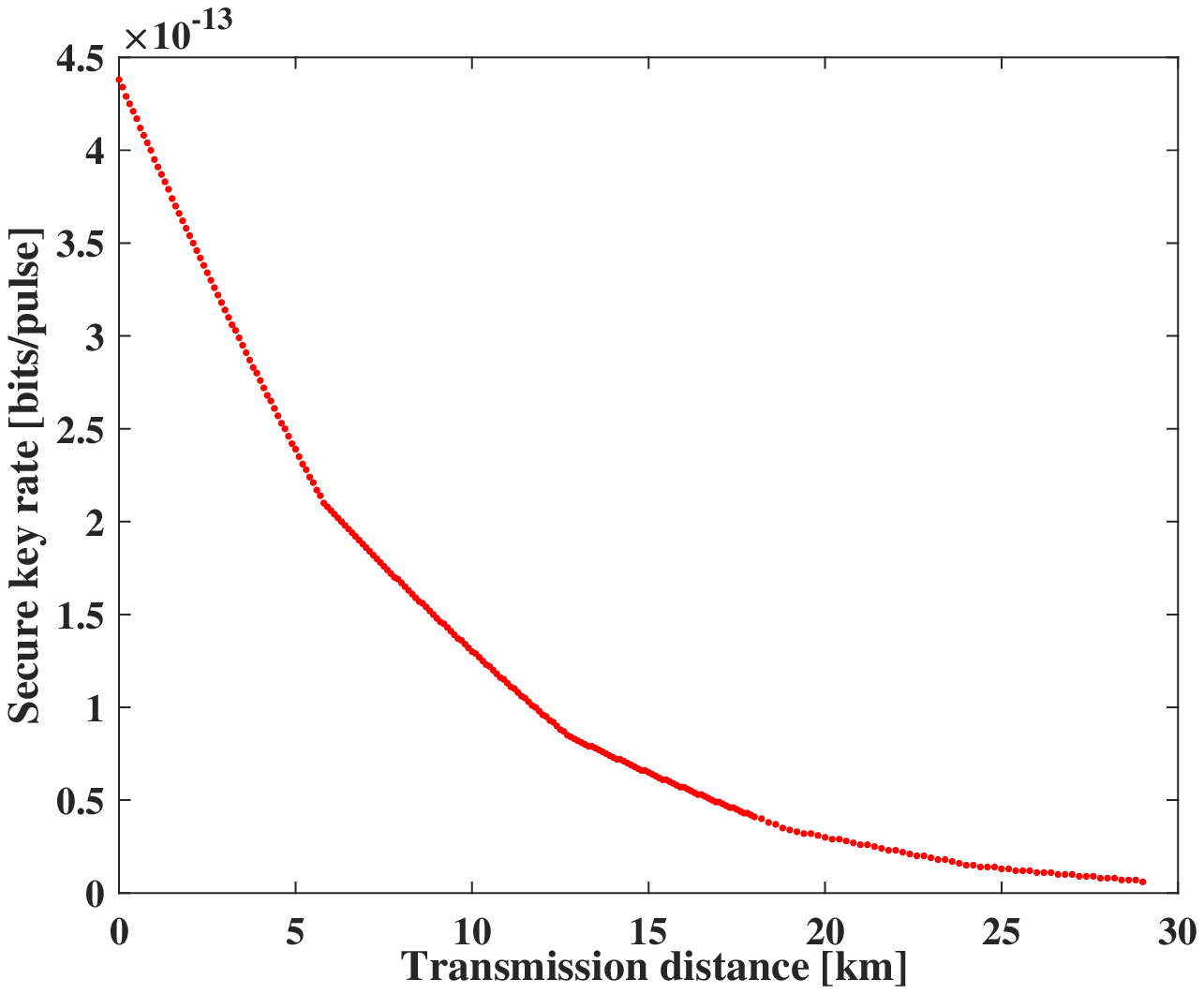}    }
	\subfigure[The relationship between the
	transmission distance and the optimal $b$ values.  ]{
		\label{fig:subfig:b}
		\includegraphics[width=3in]{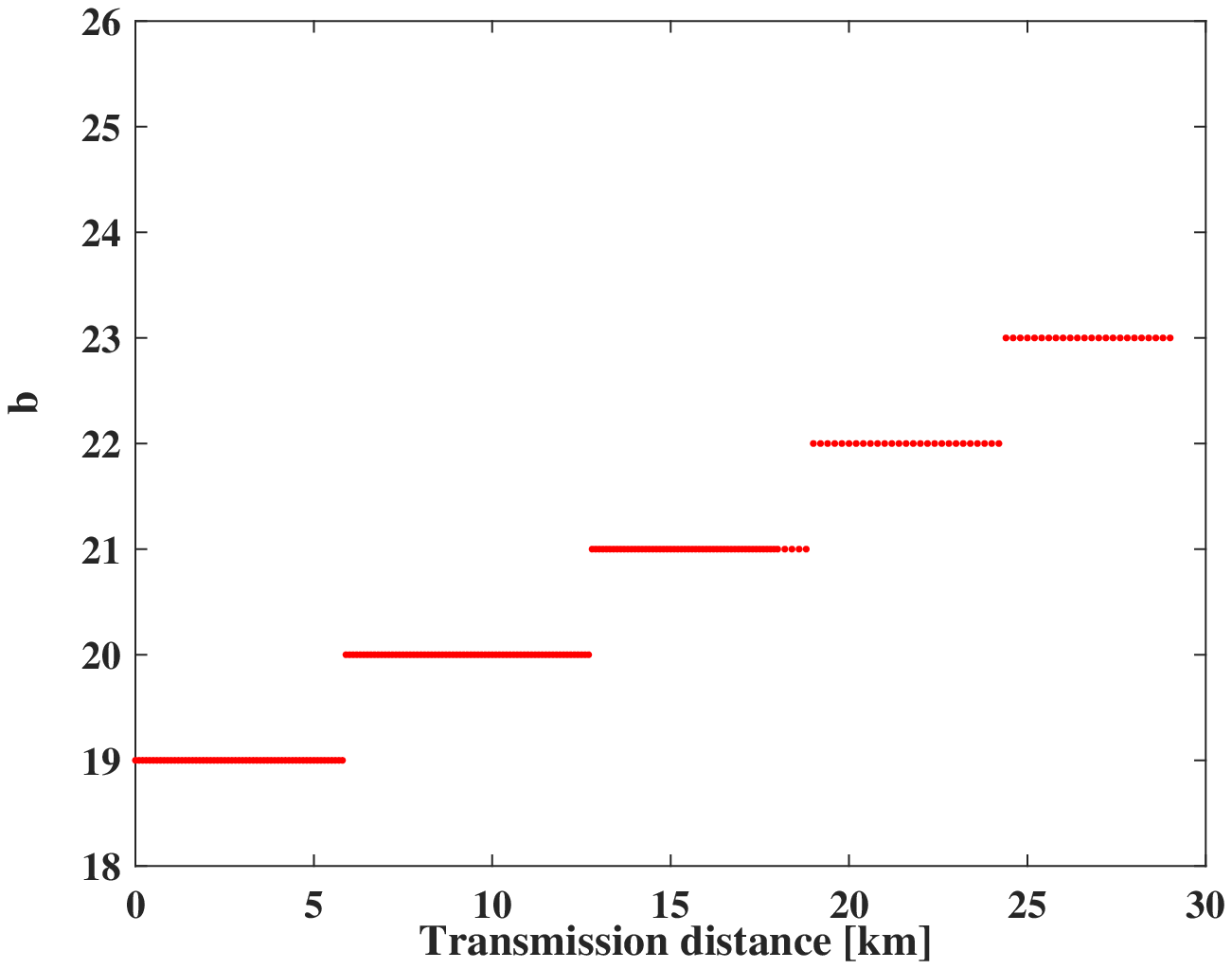}    }
	\caption{Results of TF-QKD protocol with $e_d=0.48$.}    %
	\label{fig:subfig}
\end{figure*}
More surprisingly, we find that our analysis method can still generate positive secure key, the reason for which is that $E_{\mu\mu}^{XX}$  has no correlation with $e_d$ in the security analysis model. This is quite different from the BB84-QKD system or the original MDI-QKD system, where all of the error rate in different bases will be increased by increasing $e_d$. However, in a TF-QKD system, we can prove that the error rate $E_{\mu\mu}^{XX}$  will be unchanged. By applying the AD method with $b>1$, the error rate $\tilde{E}_{uu}^{ZZ}=\frac{{E_{uu}^{ZZ}}^b}{{E_{uu}^{ZZ}}^b+(1-E_{uu}^{ZZ})^b}$ will be smaller than $E_{\mu\mu}^{ZZ}$ at the cost of increasing $\tilde{\lambda}_0$ and $\tilde{\lambda}_1$. Thus, if $b$ is large enough, we can also generate positive secure key rate $R_{TF}$ even if the error $e_d$ is close to $0.5$. From the calculation result, we find that the secure key rate is very low for high misalignment errors close to $50\%$. In this situation, the advantage distillation is very interesting from a theoretical point of view, but it is not feasible in practice.

\section{\label{sec_con}Discussion}
In a practical TF-QKD system, by combining the AD method with the information-theoretical security analysis method, we prove that both the maximal transmission distance and the secure key rate can be sharply improved. More surprisingly, the numerical simulation results demonstrate that TF-QKD can generate positive secure key even if the loss-independent misalignment error is close to $50\%$, thus our analysis method can significantly improve the
performance of a practical TF-QKD system. In the future research, it will be interesting to experimentally realize the AD method in a practical TF-QKD system, especially with a high loss-independent misalignment error.

Since the AD method only modifies the classical post-processing step, the similar advantage distillation technology has been applied in the phase-matching quantum key distribution  \cite{NJP}. In the future research, it will be also interesting  to analyze security of sending-not-sending TF QKD with advantage distillation technology  \cite{SNS1, SNS2, SNS3}. More interestingly, based on the quantum asymptotic equipartition property \cite{AEP}, the leftover hash lemma \cite{LEFTOVER} and the Chernoff bound  \cite{Chernoff, TFQKD7}, the statistical fluctuation of the bit error rate and the secure key rate in finite-key length can be efficiently analyzed in the future research.

\begin{center}{ \textbf{Acknowledgements } }\end{center}
The authors would like to thank Zhen-Qiang Yin, Rong Wang, Shuang Wang and Guan-Jie Fan-Yuan for helpful discussions. This work is supported by NSAF (Grant No. U2130205), National Natural Science Foundation of China (Grant No. 11725524, 62371244) and China Postdoctoral Science Foundation (Grant Nos. 2019T120446, 2018M642281).

\bibliographystyle{quantum}
\bibliography{apssamp}

\end{document}